\documentclass[a4paper,11pt]{article}
\usepackage{pos}
\usepackage{hyperref}
\usepackage{orcidlink}
\def\kt{\ensuremath{k_{\rm T}}}

\def\cascade{{\sc Cascade}}

\def\mcatnlo{{MCatNLO}}

\input epsf.tex
\def\desepsf(#1 width #2){\epsfxsize=#2 \epsfbox{#1}}
\def\kt{\ensuremath{k_{\rm T}}}

\def\pt{\ensuremath{p_{\rm T}}}

\def\qt{\ensuremath{q_{\rm t}}}
\def\zdyn{\ensuremath{z_{\rm dyn}}}
\def\ptll{\ensuremath{p_{\rm T}(\ell\ell)}}

\newcommand{\mdy}{\ensuremath{m_{\text{DY}}}}

\newcommand{\PBM}{PB}

\newcommand{\CAS}{{\mcatnlo+CAS3}}

\newcommand{\GeV}{\text{GeV}}

\title{NLO Analysis of Small-\boldmath\kt\ Region in Drell-Yan Production with Parton Branching}

\author*{S.~Taheri~Monfared\orcidlink{0000-0003-2988-7859}}

\affiliation{Deutsches Elektronen-Synchrotron DESY, Germany}

\emailAdd{taheri@mail.desy.de}

\abstract{The Parton-Branching Method (\PBM) facilitates the determination of Transverse Momentum Dependent (TMD) parton densities across a wide \kt\ range, spanning small to large transverse momentum scales. In the small \kt\ region, both intrinsic parton motion and resummed ultra-soft gluons are significant contributors. Our analysis highlights their crucial role in shaping integrated and TMD parton densities.

Using \PBM-derived TMD parton densities and a NLO calculation in MC@NLO style, we compute the transverse momentum spectrum of Drell-Yan pairs across a broad mass range. The spectrum's sensitivity to the intrinsic \kt\ distribution allows us to fine-tune parametric parameters. Starting from the PB-NLO-HERAI+II-2018 set2 TMD parton distributions, we determine the intrinsic \kt\ distribution width, resulting in a slightly wider profile than the default set. Importantly, this width remains independent of Drell-Yan pair mass and center-of-mass energy ($\sqrt{s}$), distinguishing our approach.
}

\FullConference{The European Physical Society Conference on High Energy Physics (EPS-HEP2023)\\
 21-25 August 2023\\
Hamburg, Germany\\}

\begin{document}

\begin{flushright}
DESY-23-181\\
\end{flushright}

\maketitle

\section{Introduction}
The Drell-Yan (DY) lepton pair production in hadron collisions serves as an essential probe of various QCD phenomena. At low transverse momentum (\pt) of the DY pair, non-perturbative parton motion within hadrons becomes significant. In this regime, resumming multiple soft gluon emissions is crucial, while at higher \pt, perturbative contributions dominate.

Various methods, including CSS, TMD resummation, and parton shower calculations, have been used to describe DY pair transverse momentum spectra across different DY masses (\mdy ) and center-of-mass energies ($\sqrt{s}$). The Parton Branching (\PBM) method ~\cite{BermudezMartinez:2018fsv,Jung:2021mox}, notably, successfully reproduces these spectra at both LHC and lower energies without parameter adjustments, in contrast to other approaches that necessitate energy-dependent intrinsic \kt\ distributions ~\cite{Gieseke:2007ad}.

Our study delves into the low \kt\ behavior of \PBM-TMD parton distributions, emphasizing the significance of including very soft gluon emissions following DGLAP. These unresolvable emissions significantly influence inclusive parton distributions, especially in the low-\kt\ TMD spectrum.
Our findings reveal that the PB-NLO-HERAI+II-2018 set2 (abbreviated as PB-set2) TMD distributions result in a minor pure intrinsic-\kt\ contribution, as most of the small-\kt\ effects are accounted for within the \PBM\ framework. 

\section{PB TMDs and DY cross section}

To investigate various low-\pt\ spectrum contributions across different \mdy\ and $\sqrt{s}$ scenarios, we employ the \PBM\ TMD method, as outlined in ~\cite{Hautmann:2017fcj,Hautmann:2017xtx}. This involves extracting NLO hard-scattering matrix elements from {\scshape MadGraph5\_aMC@NLO} and matching them with TMD parton distributions and showers derived from \PBM\ evolution, using the subtractive matching procedure introduced in ~\cite{Yang:2022qgk}.
%
%
The PB evolution equations for TMD parton distributions
$ {\cal A}_a ( x , {\bf k } , \mu^2) $
of flavor $a$ are given by:
\begin{eqnarray}
\label{evoleqforA}
   { {\cal A}}_a(x,{\bf k}, \mu^2)
 &=&
 \Delta_a (  \mu^2  ) \
 { {\cal A}}_a(x,{\bf k},\mu^2_0)
 + \sum_b
 \int {{d^2 {\bf q}^{\prime } } \over {\pi {\bf q}^{\prime 2} } }
 \
{
{\Delta_a (  \mu^2  )}
 \over
{\Delta_a (  {\bf q}^{\prime 2}
 ) }
}
\ \Theta(\mu^2-{\bf q}^{\prime 2}) \
\Theta({\bf q}^{\prime 2} - \mu^2_0)
 \nonumber\\
&\times&
\int_x^{z_M} {{dz}\over z} \;
P_{ab}^{(R)} (\alpha_s
,z)
\;{ {\cal A}}_b\left({x \over z}, {\bf k}+(1-z) {\bf q}^\prime ,
{\bf q}^{\prime 2}\right)
  \;\;  ,
\end{eqnarray}
where $z_M$ is the soft-gluon resolution scale~\cite{Hautmann:2017xtx}
, and $z$ is the longitudinal momentum transferred at the branching.
$\Delta_a$ is the Sudakov form factor.
This branching evolution fulfills soft-gluon angular
ordering and is essential for well-defined TMD distributions.

The distribution $ { {\cal A}}_a(x,{\bf k},\mu^2_0) $
at the starting scale $\mu_0$ of the evolution is a nonperturbative boundary condition to the evolution equation, and is
to be determined from experimental data. We parameterize it as:
\begin{equation}
\label{TMD_A0}
{\cal A}_{0,b} (x, k_T^2,\mu_0^2)   =  f_{0,b} (x,\mu_0^2)
\cdot \exp\left(-| k_T^2 | / 2 \sigma^2\right) / ( 2 \pi \sigma^2) \; ,
\end{equation}
with $ \sigma = q_s / \sqrt{2} $,
independent of parton flavor and $x$, where $q_s$ is the intrinsic-$k_T$ parameter.
The scale at which the strong coupling $\alpha_s$ is to be evaluated in
Eq.~(\ref{evoleqforA}) can follow two scenarios, either $\alpha_s ({\bf q}^{\prime 2})$ or $\alpha_s (q_T^2)$ while $q_T^2={\bf q}^{\prime 2} (1-z)^2$.
In this study, we focus on the scenario where $\alpha_s$ is determined as a function of the transverse momentum $q_T^2$, referred to as PB-set2. This scenario provides a better description of various measurements and utilizes an intrinsic-$k_T$ parameter $q_s = $ 0.5 GeV.



\section{Non-perturbative contribution}
The PB TMD method incorporates Sudakov evolution through phase space integrations of relevant kernels over the resolvable region, encompassing momentum transfers $z$ up to the soft-gluon resolution scale $z_M$. 
The effects of $z_M$ on parton distributions and transverse momentum distributions are significant. The choice of $z_M$ mainly affects the soft region, while the perturbative region with $\qt > q_0$ remains unaffected by this choice. A scale like $z_M=\zdyn = 1 - q_0/\mu^\prime$ removes emissions with $\qt < q_0$. In contrast, using $z_M \to 1$ includes very soft emissions by default. The intrinsic-\kt\ distribution has a more pronounced effect at smaller scales, but its contribution decreases at larger scales. 
The detailed description of the non-perturbative contribution for PB-set1 and PB-set2 can be found in Ref.~\cite{soft-gluon,intkt}.
\section{DY Cross Section Computation}
The cross section for DY production is computed at NLO using the MCatNLO method. In this approach, the collinear and soft contributions from the NLO cross section are subtracted, as they will be included later when applying parton showers or TMD parton densities. We incorporate TMD parton distributions and parton showers into the MCatNLO calculation using \cascade 3 \cite{Baranov:2021uol}, as discussed in detail in Ref. \cite{Yang:2022qgk}. We utilize the Herwig~6 subtraction terms in MCatNLO, which are consistent with the PB-set TMD parton distribution sets described earlier.
 The predicted cross sections are calculated using the integrated versions of the NLO parton densities PB-set, along with $\alpha_s(m_Z) = 0.118$ at NLO.

The factorization scale $\mu$ for the hard process calculation is set to $\mu = \frac{1}{2} \sum_i \sqrt{m_i^2 + p_{t,i}^2}$, where the sum includes all final state particles. For DY production, this includes all decay leptons and the final jet. When generating transverse momentum according to the PB distributions, $\mu$ is set to $\mu = \mdy$. In the case of real emission, it is set to $\mu = \frac{1}{2} \sum_i \sqrt{m_i^2 + p_{t,i}^2}$. The generated transverse momentum is constrained by the matching scale $\mu_m = \text{SCALUP}$ \cite{Baranov:2021uol}.
As there are currently no \PBM-fragmentation functions available, the final state parton shower in \cascade 3 is generated using Pythia, including photon radiation from the lepton pair.

\section{The transverse momentum distribution of Drell-Yan lepton pairs}
The transverse momentum spectrum of DY lepton pairs at $\sqrt{s}=13$TeV has been measured across a wide range of $\mdy$ values ($\mdy=[50, 76, 106, 170, 350,
1000]$~GeV) by CMS~\cite{CMS:2022ubq}. This measurement is provided including a detailed uncertainty breakdown, including a complete treatement of experimental uncertainties with correlations between bins of the measurement.

We compare these measurements with predictions from \CAS, utilizing PB-sets. Notably, as previously observed in various studies \cite{Yang:2022qgk,Abdulhamid:2021xtt,Martinez:2020fzs,BermudezMartinez:2019anj}, PB-set1 tends to overestimate contributions at low transverse momenta (\ptll), while PB-set2 aligns well with the measurements without requiring additional parameter adjustments. Given the success of \CAS\ with PB-set2 in describing the DY \ptll-spectrum in the low \ptll-region, we now explore the significance of the intrinsic-\kt\ distribution. In PB-sets, this distribution is represented as a Gaussian as shown in Eq.~\ref{TMD_A0}. 
We focus on \ptll\ values below the peak region to avoid the absence of higher-order contributions in the matrix element.
We use the detailed breakdown of the experimental uncertainties provided on the CMS public website.
To determine of the intrinsic-\kt\ we vary the $q_s$ parameter and calculate a $\chi^2$ to quantify the model agreement with the measurement. Detailed discussion on the treatment of the uncertainties and calculation of $\chi^2$ is available in Ref. ~\cite{intkt}.

\begin{figure}
    \centering
    \includegraphics[width=0.49\textwidth]{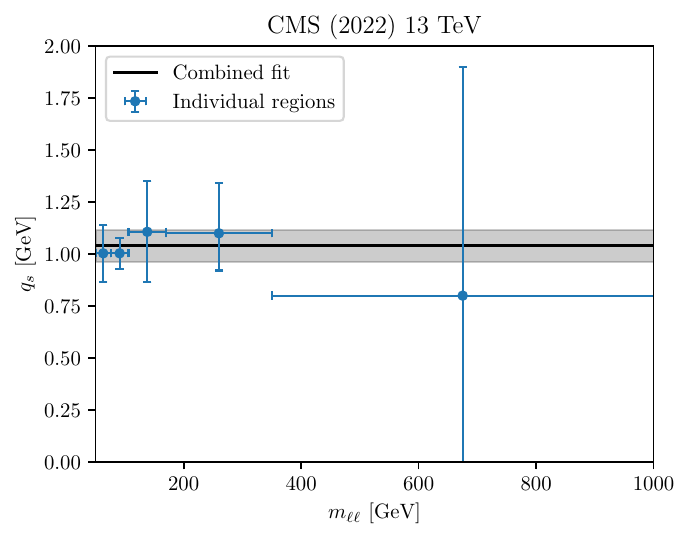}
    \caption{%
        The values of $q_s$ obtained in each \mdy -bin as obtained from 
        Ref.~\protect\cite{CMS:2022ubq}. Indicated is also the final value of $q_s$.
    }
    \label{fig:qs_vrs_mdy-13TeV}
\end{figure}
\begin{figure}[h!tb]
\begin{center} 
\includegraphics[width=0.49\textwidth]{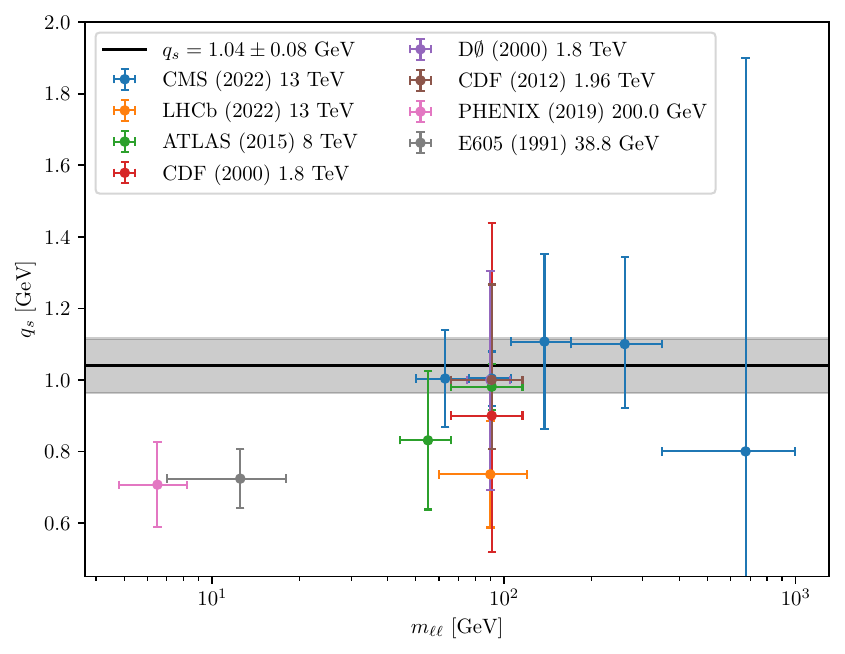}
\includegraphics[width=0.49\textwidth]{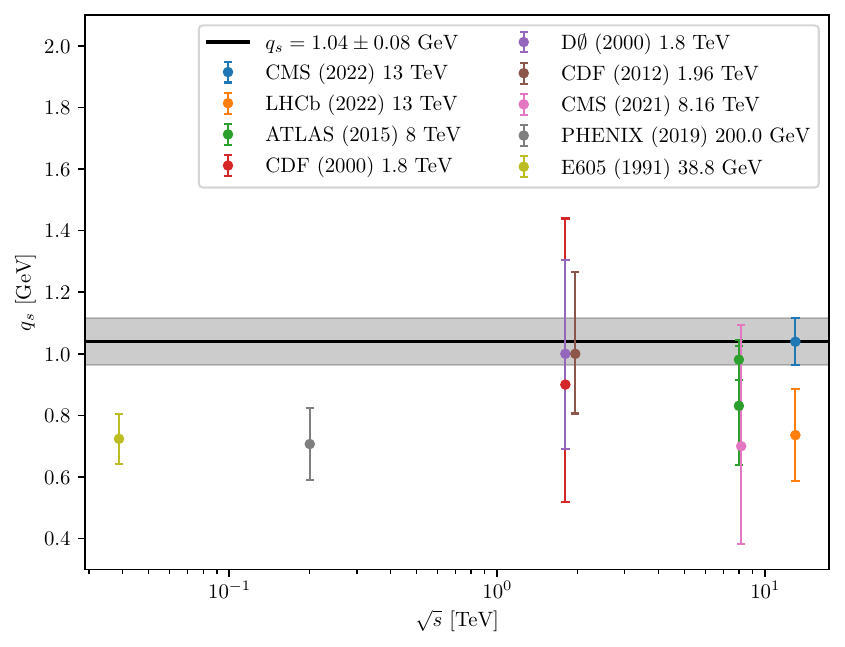}
 \caption{\small  
{\bf Left}: the value of $q_s$ as a function of the DY-mass as obtained from the measurements in 
  Refs.~\protect\cite{CMS:2022ubq,LHCb:2021huf,CMS:2021ynu,Aad:2015auj,CDF:1999bpw,D0:1999jba,CDF:2012brb,Aidala:2018ajl,Moreno:1990sf}. 
  {\bf Right} : same as a function of $\sqrt{s}$.
 }
\label{qsversusmass}
\end{center}
\end{figure}

The best fit value, extracted using the correlated uncertainties, is: $q_s = 1.04\pm0.03(\text{data})\pm0.05(\text{scan})\pm0.05(\text{binning})\:\GeV$.
This value and its uncertainty are shown as a black line and shaded area in
Fig.~\ref{fig:qs_vrs_mdy-13TeV} for comparison with the individual \mdy\ bins.
%
Figure~\ref{qsversusmass} illustrates the variation of $q_s$ as a function of \mdy\ and $\sqrt{s}$, utilizing data from different measurements mentioned in Refs.\protect\cite{CMS:2022ubq,LHCb:2021huf,CMS:2021ynu,Aad:2015auj,CDF:1999bpw,D0:1999jba,CDF:2012brb,Aidala:2018ajl,Moreno:1990sf}. Notably, the value of $q_s=1.04 \pm 0.08$\GeV, determined from the measurements in Ref.\cite{CMS:2022ubq}, remains applicable across all \mdy\ ranges and for various $\sqrt{s}$ values.

\section{Summary and Conclusion}
We here apply the PB approach to low \kt\ DY production, presenting
results from the work in Ref. \cite{intkt}.
In this study, we discuss the \PBM-TMD distributions, focusing on soft and low transverse momenta contributions. We obtain the NLO cross section of the DY process and compare it to recent LHC measurements. We emphasize the importance of the soft, non-perturbative region for both integrated and transverse momentum distributions.
The key result of this work is the extraction of the intrinsic-$k_T$ parameter $q_s$ from measured \pt\ dependence of DY cross sections at various $\sqrt{s}$ and \mdy . We find a consistent value of $q_s = 1.04 \pm 0.08$ \GeV , independent of \mdy\ and $\sqrt{s}$, in agreement with expectations from Fermi motion in protons. This contrasts with typical Monte Carlo generators that require $\sqrt{s}$ and \mdy-dependent widths for intrinsic Gauss distributions.
The stability of our obtained $q_s$ value is influenced by the "non-perturbative Sudakov form factor," which is crucial for stable integrated distributions.

\section*{Acknowledgments}
The results discussed in this contribution are based on the work in Ref. \cite{intkt}. Many thanks to all co-authors for collaboration. I am grateful to the organizers of EPS2023 for the invitation to present these results at the workshop.

\end{document}